\documentclass[12pt]{JHEP3} 

\usepackage{epsfig,multicol,rotating,axodraw,epsfig,subfigure}

\newcommand\fverb{\setbox\pippobox=\hbox\bgroup\verb}
\newcommand\fverbdo{\egroup\medskip\noindent%
                        \fbox{\unhbox\pippobox}\ }
\newcommand\fverbit{\egroup\item[\fbox{\unhbox\pippobox}]}
\newbox\pippobox

\title{$B_s \rightarrow \mu^+ \mu^-$ and  the upward-going
muon flux from the WIMP annihilation 
in the sun or the earth}

\author{Seungwon Baek \\
        Laboratoire Ren\'e J.-A. L\'evesque, Universit\'e de Montr\'eal,
        C.P. 6128, succ. centre-ville, Montr\'eal, QC, Canada H3C 3J7\\
        E-mail: \email{}}
\author{Yeong Gyun Kim \\
       Department of Physics, Korea University, Seoul 136--701, Korea \\
       E-mail: \email{yg-kim@korea.ac.kr}}   
\author{P. Ko \\
        School of Physics, KIAS, Seoul 130--722, Korea \\
        E-mail: \email{pko@kias.re.kr}} 

\preprint{KIAS-P05029}      

\abstract{
We consider the upward-going muon flux due to the WIMP annihilations 
in the cores of the sun and the earth, including the upper bound 
on the branching ratio for $B_s \rightarrow \mu^+ \mu^-$ decay. 
We find that the constraint from $B_s \rightarrow \mu^+ \mu^-$ is 
very strong in most parameter space, and exclude the supergravity 
parameter space regions where the expected upward-going muon fluxes
are within the expected reach of AMANDA II. 
}


\dedicated{}

\begin{document} 


\section{Introduction}
\label{sec:introduction}
There is now compelling evidence for a non-baryonic cold dark matter 
(DM) component  in the universe \cite{Spergel:2003cb}.
In the Minimal Supersymmetric Standard Model (MSSM) with $R$ parity,
the lightest supersymmetric particle (LSP) is stable and becomes 
a good candidate for cold dark matter in the universe \cite{lspdm}. 
The LSP is often the lightest neutralino which is the admixture of
Bino, Wino and Higgsinos in the MSSM. In this case, the neutralino DM 
in our galactic halo might be detected via its elastic scattering 
with terrestrial nuclear targets \cite{Goodman,Jungman}.
In fact, the DAMA Collaboration \cite{dama} even claimed an evidence for DM.
However, the CDMS II experiment \cite{cdms} has reported the upper limit
on the DM scattering cross section, which is not compatible with 
the results of the DAMA experiment. There are several experiments going 
on searching for the DM scattering at the level of 
$\sigma_{\chi p}^{\rm SI} \sim 10^{-7}$ pb or less.
In the most widely studied minimal supergravity (mSUGRA) scenario 
(or the constrained MSSM), the spin-independent neutralino-proton
 scattering cross section
$\sigma_{\chi p}$ turns out very small ($\lesssim 10^{-8}$ pb). 
However there is no solid theoretical rationale for the minimal supergravity
scenario, and it is important to calculate the possible maximal values for
$\sigma_{\chi p}$ in {\it general} supergravity scenarios 
beyond the mSUGRA scenario.
And it is very important to impose all the relevant constraints from 
various experiments in order not to overestimate the cross section.
Some important constraints include the lower bounds on the Higgs and SUSY 
particle masses, $B\rightarrow X_s \gamma$ branching ratio, the muon 
$(g-2)_\mu$, etc.. One may also take some theoretical consideration 
on the absence of the color-chrage breaking minima or the directions 
unbounded from below, etc..

In a previous work \cite{bsdm1}, we pointed out that there is a strong 
correlation between the spin independent neutralino-proton scattering 
cross section $\sigma_{\chi p}$ and the branching ratio for 
$B_s \rightarrow \mu^+ \mu^-$ decay \cite{babu,kane,Baek:2002rt,Baek:2002wm,ellis} 
within mSUGRA and its extensions. 
The origin of this correlation resides in the dependence of both observables
on $\tan\beta$ and the neutral Higgs boson masses; both observables increase
for large $\tan\beta$ and low Higgs masses. In particular, we have shown that
the current upper limit on $B(B_s \rightarrow \mu^+ \mu^-)$ excludes 
substantial parameter space where the DM scattering cross section is within 
the CDMS sensitivity region \cite{bsdm1} 
(see also \cite{bckkm} for a detailed analysis).

In this work, we extend our previous study to the indirect detection of 
neutralino DM with neutrino telescope through upward-going muon flux, 
and its correlation with $B(B_s \rightarrow \mu^+ \mu^-)$.
The energetic neutrino(-induced muon) flux from neutralino DM annihilation 
in the sun and the earth is one of the promising signals in the  indirect 
detection of neutralino DM \cite{indirect}. 
Neutralino DM particles in the halo can be captured by the sun or by 
the earth, when their velocities drop below escape velocities 
via their elastic scattering with matter in the sun or earth.  
Then they will accumulate in the core of the sun and the 
earth and will eventually annihilate into ordinary SM particles. 
Among the annihilaion products, neutrinos can pass through the sun and the 
earth, and then could be detected in neutrino telescopes through their 
conversion to muons via charged-current scattering with neuclei near 
the detectors. 
Baksan \cite{baksan}, MACRO \cite{macro}, Super-K \cite{superk} and 
AMANDA \cite{amanda} released upper limits on the upward-going 
muon flux. There are also planned or proposed neutrino telescopes such as
ANTARES \cite{antares}, IceCube \cite{icecube} and NESTOR \cite{nestor} etc..

An important point of the indirect detection of neutralino DM with 
neutrino telescopes is that the neutrino flux strongly depends on the 
capture rate of neutralino by the sun or the earth, which in turn 
depends on  neutralino-nucleon scattering cross sections.
Therefore we expect some correlation between the neutrino flux and 
$B(B_s \rightarrow \mu^+ \mu^-)$, which is similar to the strong correlation
between $\sigma_{\chi p}$ and $B(B_s \rightarrow \mu^+ \mu^-)$ as discussed
in Ref.~\cite{bsdm1}.
Indeed, we will show that the current upper limit of 
$B(B_s \rightarrow \mu^+ \mu^-) < 4.1 \times 10^{-7}$ (90 \% CL) 
\cite{cdf,d0} puts  
strong constraints on the 
upward-going muon flux in the supersymmetric models which give rather 
large spin-independent neutralino-proton scattering cross section. 

This paper is organized as following. In Sec.~~2, we give a brief review
on the indirect detection of the DM through the upward-going muon flux.
In Sec.~3, we consider the upward-going muon fluxes in some supergravity
scenarios and illuminate our point that $B_s \rightarrow \mu^+ \mu^-$ 
branching ratio plays an important role. In Sec.~4, we summarize the
results. 

\section{Indirect detection through the upward-going muon flux}
\label{sec:indirect}

As we mentioned in the introdecution, the observation of energetic 
neutrinos from the sun and/or the earth would 
provide convincing evidence of the existence of neutralino dark matter 
in galactic halo \cite{Jungman}.  The flux of energetic neutrinos from neutralino 
annihilation in the sun or the earth is proportional to the rate of 
neutralino annihilation in the sun or in the earth and the 
energy spectrum of neutrinos from the annihilation.  
The time evolution of the number of neutralino, $N$ in the sun 
(or in the earth) is given by
\begin{eqnarray}
\dot{N} = C - C_A N^2
\end{eqnarray}
where $C$ is the capture rate of neutralino by the sun or the earth and 
$C_A$ is the total annihilation cross section times relative velocity 
per volume.  From Eq.(2.1), we find that the present annihilation rate is 
\begin{eqnarray}
\Gamma_A = \frac{1}{2} C_A N^2 = \frac{1}{2} C \tanh^2 (\sqrt{C C_A}~ t_0)
\end{eqnarray}
where $t_0 \simeq 4.5$ Gyr is the age of the solar system.
For $\sqrt{C C_A}~t_0 \ll 1$, the annihilation rate is 
$\Gamma_A \approx \frac{1}{2} C^2 C_A t_0^2$ and less than its maximal value.
But, for $\sqrt{C C_A}~t_0 \gg 1$, the neutralino density reach equilibrium
and the annihilation rate is $\Gamma_A \approx \frac{1}{2} C$.
Therefore, when accretion is efficient, the annihilation rate depends on the
capture rate $C$, but not on the annihilation cross section.

In turn, the capture rate $C$ strongly depends on the elastic scattering cross section of 
neutralino with matter in the sun and the earth. 
The capture rate for the earth primarily depends on the spin-independent 
DM scattering  cross section.  For the capture rate in the sun, however, 
both spin-independent and spin-dependent scattering cross section can be 
important and the significance of
each contribution depends on the specific SUSY scenarios.

In MSSM, $t$-channel Higgs boson and $s$-channel squark exchange processes
contribute to the spin-independent (scalar) scattering between neutralino and quarks.
In many case, dominant contribution to the scalar cross section 
comes from the Higgs exchange process, 
which increases for large $\tan\beta$ and small Higgs masses
and also if neutralino is a mixed gaugino-Higgsino state.
On the other hand, for the spin-dependent cross section, $t$-chennel $Z$ boson and 
$s$-chennel squark exchange processes contribute. 
Usually $Z$ exchange contribution dominates, which
is sensitive to Higgsino components of LSP,
but largely independent of $\tan\beta$. Note that if the Higgsino component 
of the LSP increases, then both the spin-independent and the spin-dependent
scattering cross sections will be enhanced, as shown below in the 
nonuniversal Higgs mass parameter case.

The capture rate $C$ also depends on the local density of neutralino, $\rho_\chi$ 
and the neutralino velocity dispersion in the halo, $\bar{v}$ etc. 
For our numerical calculation, we use the code DARKSUSY \cite{darksusy} 
and fix $\bar{v} = 270~ \rm{km/s} $. For the local density of neutralino 
we fix $\rho_\chi = 0.3~ \rm{GeV/cm^3}$ if $\Omega_\chi h^2 \geq 0.025$, while performing
a rescaling of the density as $\rho_\chi \rightarrow \rho_\chi~ (\Omega_\chi h^2 / 0.025)$
if $\Omega_\chi h^2 < 0.025$. Here $\Omega_\chi$ is the neutralino relic density
in units of the critical density and $h$ is the present Hubble constant in units
of $100~ \rm{km~s^{-1}~Mpc^{-1}}$.
\section{Upward-going muon flux in SUSY models}
\label{sec:models}

%
\subsection{mSUGRA}
In mSUGRA model, we assume a universal SUSY breaking
scalar mass $m$, a universal gaugino mass $M$ and a universal trilinear 
coupling $A$  at GUT scale $m_{\rm GUT} \simeq 2 \times 10^{16}$ GeV.  
We also require that electroweak symmetry break radiatively and then 
the Higgsino mass parameter $\mu$ is determined by the condition:  
\begin{equation}
\mu^2 = {{m_{H_d}^2 - m_{H_u}^2 \tan^2 \beta } \over {\tan^2 \beta - 1 }}
- {1\over 2} M_Z^2 .
\end{equation}
where $\tan\beta$ is the ratio of two Higgs vacuum expectation values and
$m_{H_u}^2, m_{H_d}^2$ are the soft breaking Higgs masses-squared. 
With the above mSUGRA assumptions, $|\mu|$ is usually large so that 
the lightest neutralino is bino-like and the pseudo-scalar Higgs mass 
$m_A$ is rather large. 

\begin{figure}[ht!]
\begin{center}
\subfigure[$\Phi_\mu$ (sun)]{
\includegraphics[height=7cm,width=7cm]{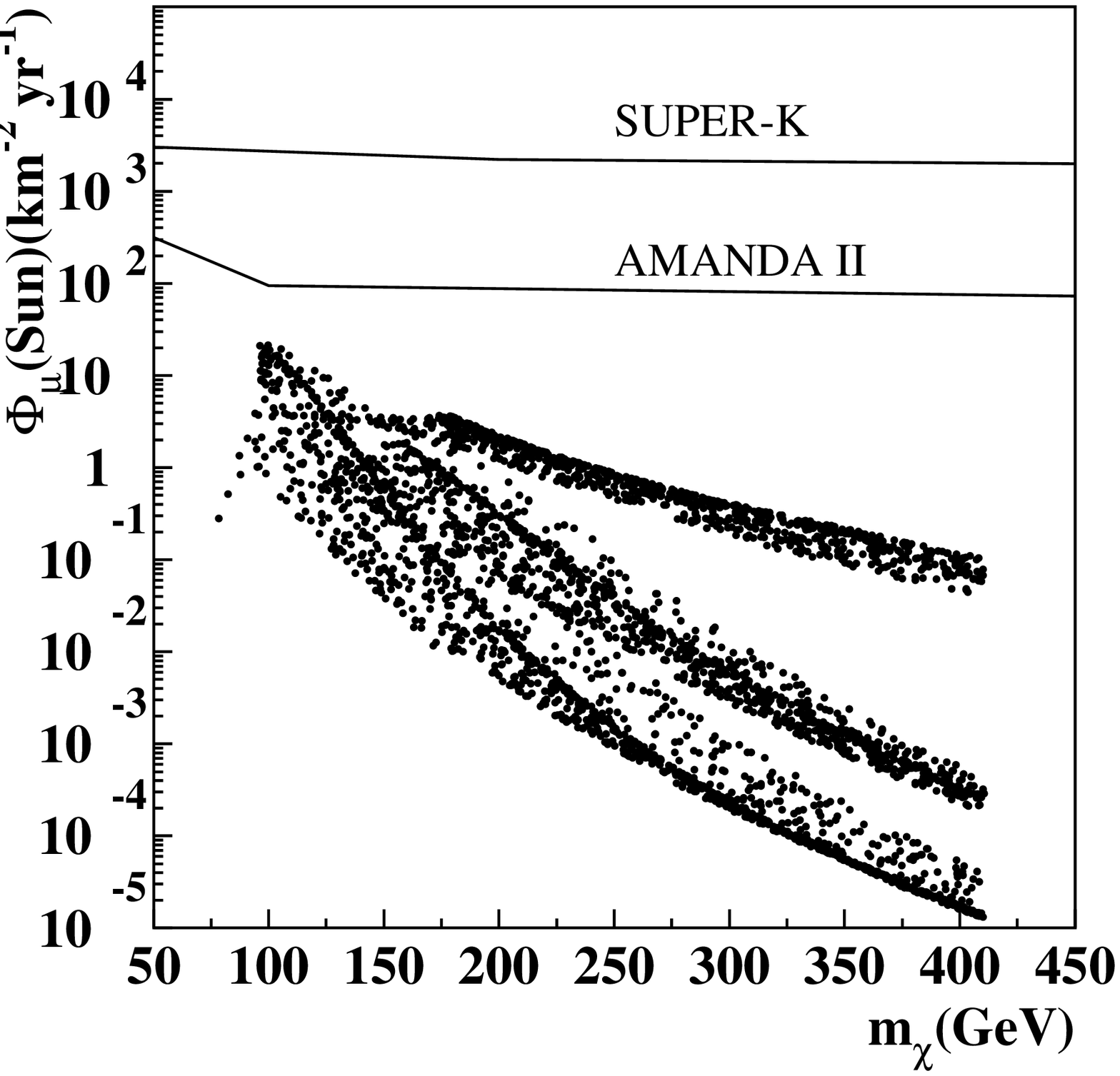}}
\subfigure[$\Phi_\mu$ (earth)]{
\includegraphics[height=7cm,width=7cm]{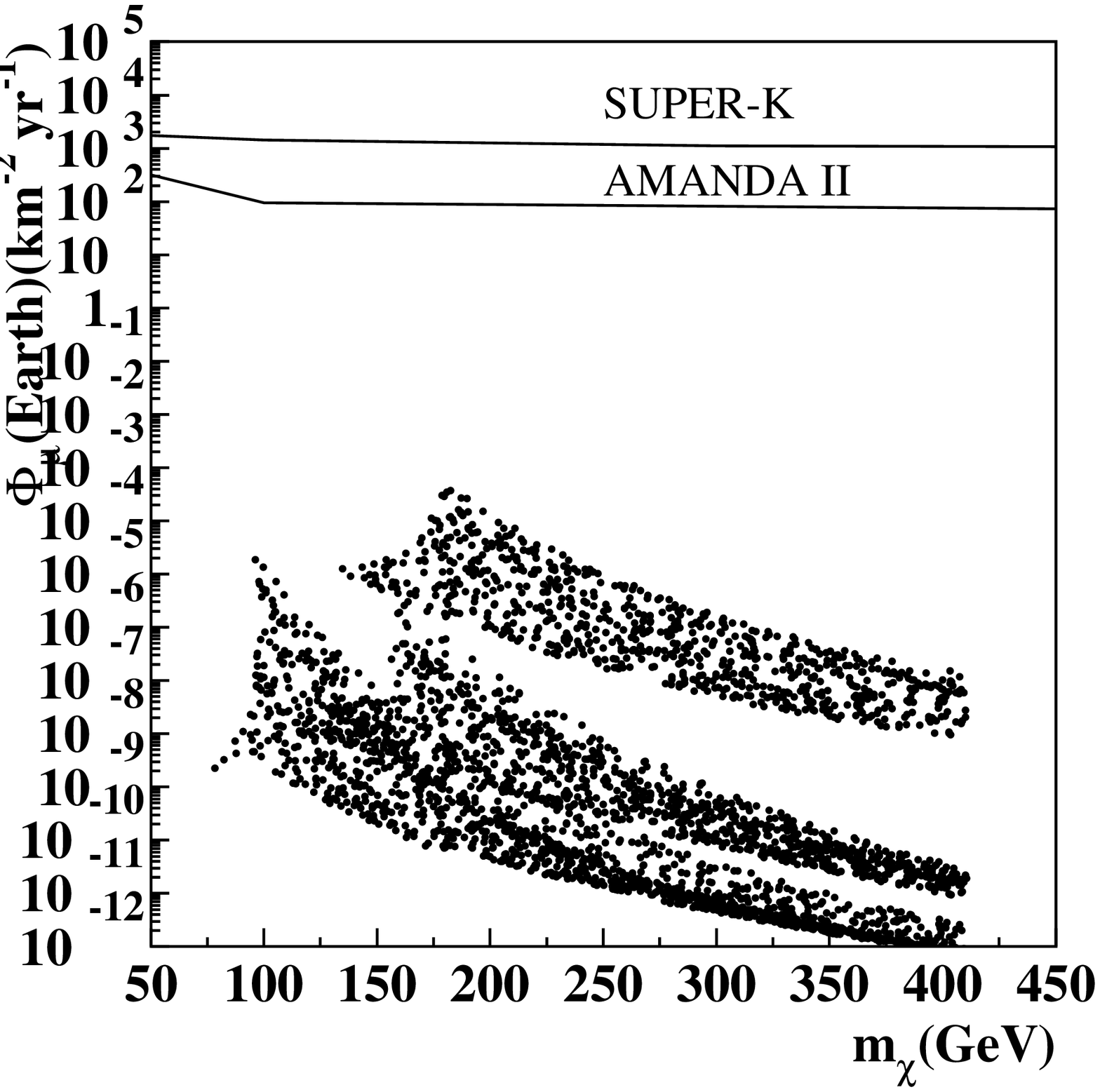}}
\end{center}
\vskip -0.5cm
\caption{The muon flux from the sun and the earth  
vs. $m_\chi$ in mSUGRA scenarios. The three branches correspond to
$\tan\beta=$10,35 and 50 respectively (from bottom to top).
The current upper limit from SUPER-K and the expected reach \cite{baltz}
of AMANDA II are also illustrated.}
\label{fig:msugra}
\end{figure}

In Fig.s~\ref{fig:msugra} (a) and (b), we show the allowed ranges of 
the upward-going muon fluxes from the sun and the earth respectively 
as  functions of the LSP mass. 
The three branches correspond to $\tan\beta=10,35$ and 50 cases 
(from the bottom to the top), respectively. 
Here, we took $A=0$ and $\mu > 0$ (motivated by 
the muon $(g-2)_\mu$ experiment) and varied  $m$ and $M$  up to 1 TeV.
We have imposed the experimental bounds for the Higgs and sparticle masses 
and for $b \rightarrow s\gamma$ branching ratio. We also required that 
the lightest neutralino is LSP. For opposite sign of $\mu$, the muon flux 
could be smaller. But we are interested in the possible maximal values, 
and we consider the positive $\mu$ case only in this work. 

In our scan, the muon flux from the sun reaches up to 
$\sim 20~ km^{-2} yr^{-1}$ if the neutralino LSP is light enough 
($m_\chi \sim 100$ GeV). 
In the small $m_\chi$ region, the neutralino density in the sun
can reach (near) equilibrium so that the muon flux is more or less 
determined by the capture rate of neutralino by the sun. 
And in turn, the capture rate in the sun is determined primarily 
by the spin-dependent scattering cross section. 
(though the contribution from the spin-independent scattering cross section 
to the capture rate can be comparable to the one from the spin-dependent 
cross section  for very large $\tan\beta$ cases)
As we already mentioned in Sec. 2, the spin-dependent scattering cross 
section is largely independent of $\tan\beta$. 
Therefore the upward-going muon flux from the sun in the small $m_\chi$ 
region gives similar values for the three choices of $\tan\beta$ values,
as one can check from the figure. 

In large $m_\chi$ region, the neutralino density is usually far from
equilibrium since the elastic scattering cross section of the DM with 
ordinary matter in the sun becomes smaller, 
and the neutralino annhilation cross section becomes important for 
the prediction of the muon flux.
An important process in this case is the neutralino pair annihilation into 
$b\bar{b}$  through s-channel pseudo-scalar Higgs exchange diagram, 
which is strongly enhanced for large $\tan\beta$ \cite{Nezri}. 
From the Fig.~\ref{fig:msugra} (a), one can notice a clear dependence of 
the muon flux  from the sun on $\tan\beta$ in the large $m_\chi$ region.

For the muon flux from the earth, the neutralino density is far less than 
the equilibrium  values and both the capture rate and the annhilation rate 
are important for the calculation  of the muon flux.  
The resulting flux is much below the one from the sun.
The maximal value of the muon flux from the earth is about 
$3 \times 10^{-5}~ km^{-2} yr^{-1}$, which is far below the SUPER-K and 
AMANDA II  sensitivity regions.

Note that there is no further constraint from the 
$B_s \rightarrow \mu^+ \mu^-$ bound for the mSUGRA case, once we impose 
the constraints from the lower bounds for Higgs boson and SUSY particle 
masses and the $B\rightarrow X_s \gamma$ branching ratio, as discussed in 
Ref.~\cite{bsdm1}. 

\subsection{Non-universal Higgs model (NUHM)}

In the previous subsection, we have shown that the mSUGRA assumption 
predicts the muon fluxes from the sun and the earth that are far below  
the sensitivity region of the current experiments.
This is mainly because the lightest neutralino is bino-like and 
the pseudo-scalar Higgs mass is large in the scanned region of mSUGRA 
scenario.  Larger muon flux from the sun and the earth can be obtained 
if we relax the universal boundary condition at GUT scale.

In this subsection we consider the non-universal Higgs model,
in which the assumption of universal soft scalar masses 
are relaxed for soft Higgs masses, as follows:
\begin{equation}
m_{H_u}^2 = m^2 ~( 1 + \delta_{H_u} ),~~~
m_{H_d}^2 = m^2 ~( 1 + \delta_{H_d} ),~~~
\end{equation}
whereas other scalar masses still have a universal mass $m$ at GUT scale.
Here $\delta$'s are parameters with $\lesssim O(1)$.

\begin{figure}[ht!]
\begin{center}
\subfigure[$\tan\beta = 35$]{
\includegraphics[height=7cm,width=7cm]{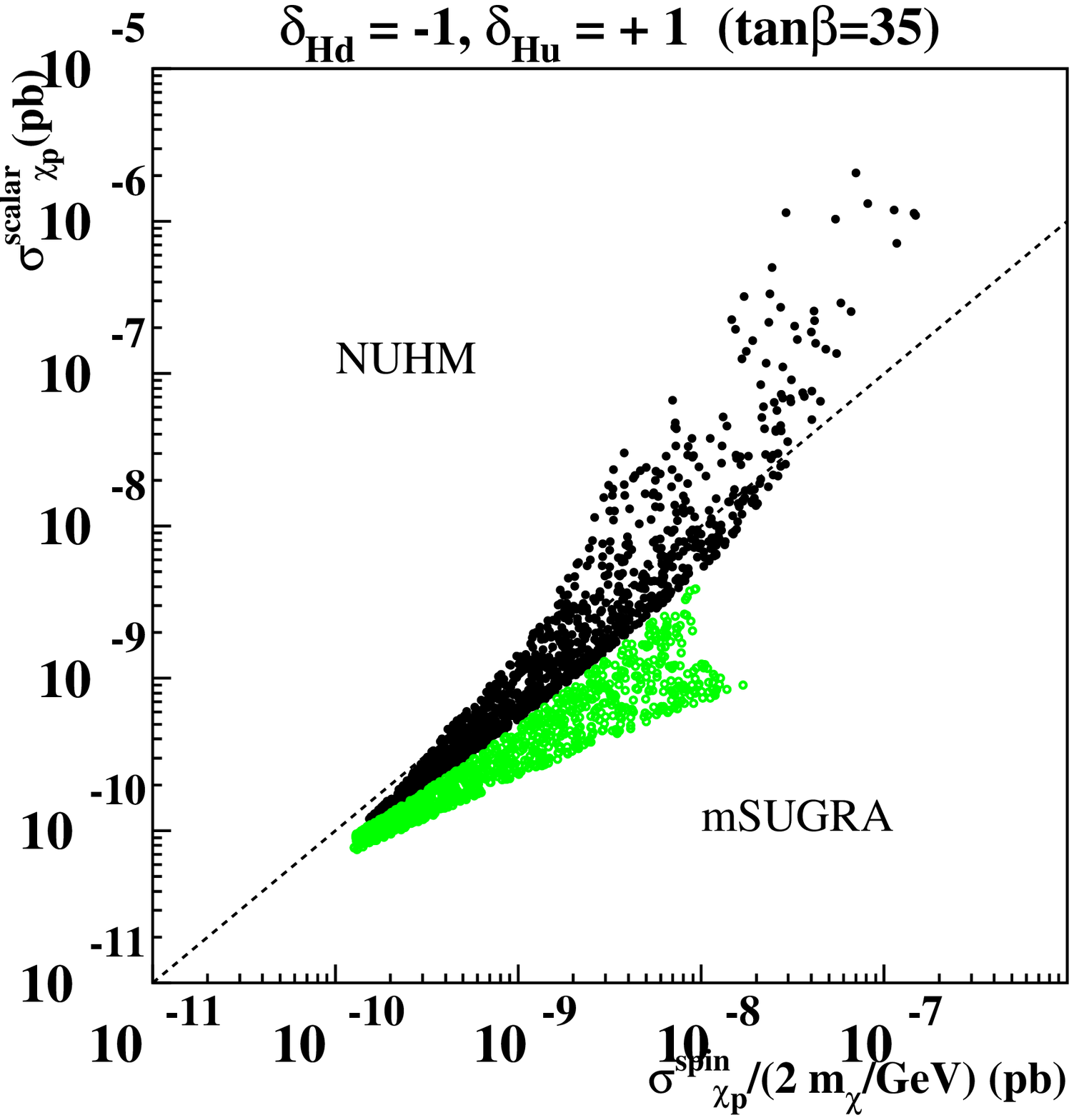}}
\subfigure[$\tan\beta = 50$]{
\includegraphics[height=7cm,width=7cm]{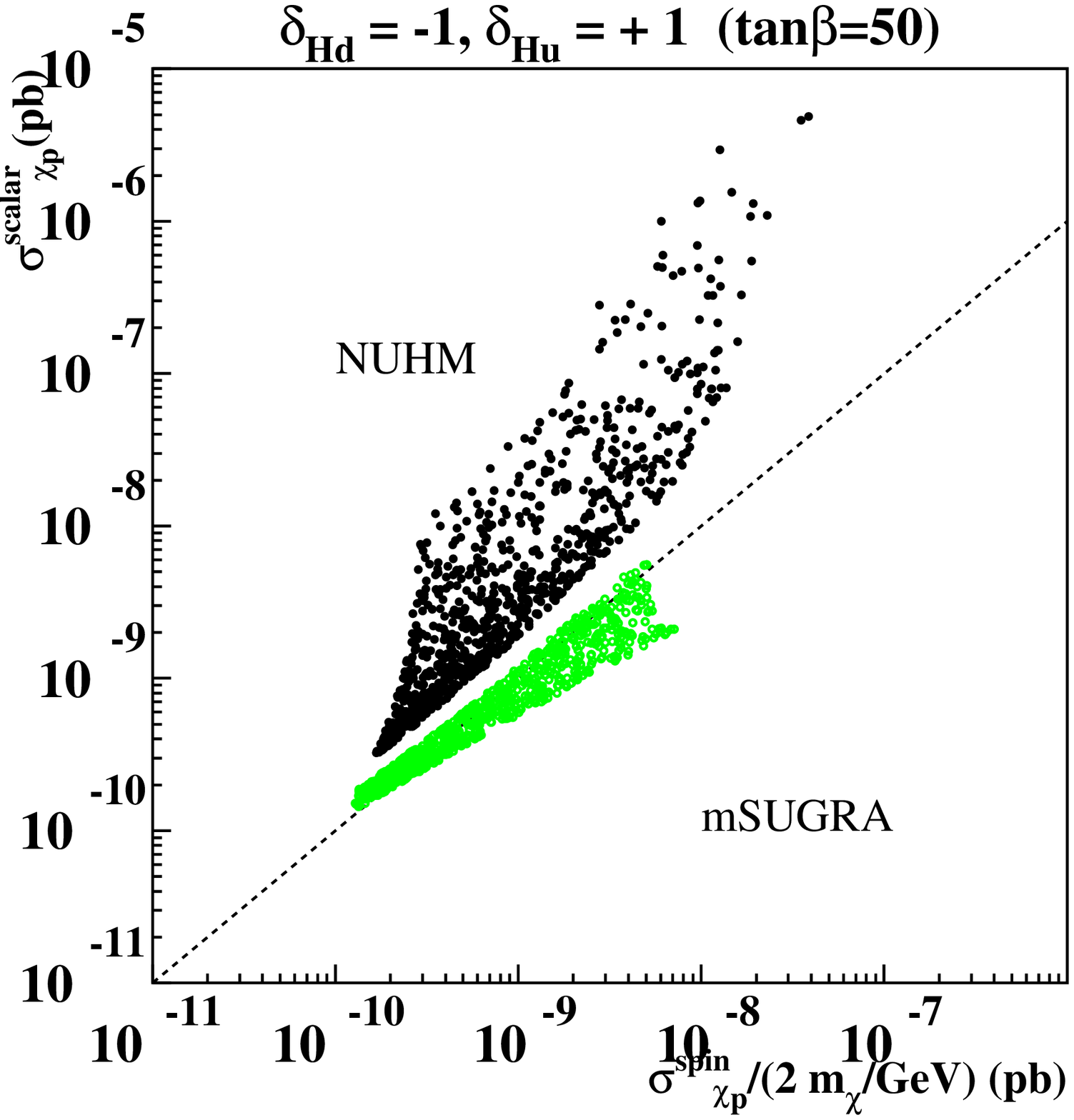}}
\end{center}
\vskip -0.5cm
\caption{$\sigma^{scalar}_{\chi p}$ vs. 
$\sigma^{spin}_{\chi p} /(2 m_\chi/GeV)$
in NUHM (black) and in mSUGRA (green) for (a) $\tan\beta=35$ and 
(b) $\tan\beta = 50$. }
\label{fig:sigsip-sigsdp-tb35}
\end{figure}

As an optimal choice for enhancing the muon flux from the sun
and the earth, we take the numerical values of $\delta'$s
as $\delta_{H_d}=-1$ and $\delta_{H_u}=1$. 
For the postive $\delta_{H_u}$, $\mu$ becomes lower and the Higgsino component
in the neutralino LSP increases so that $\sigma_{\chi p}$ is enhanced, as
discussed in Ref.~\cite{Munoz:2003wx,Cerdeno:2003yt}.
The change of $|\mu|$ also has an impact on the Higgs masses because
\[
m_A^2 = m_{H_u}^2 + m_{H_d}^2 + 2\mu^2 \simeq m_{H_d}^2 + \mu^2 - M_Z^2/2
\]
at weak scale.  For the negative $\delta_{H_d}$, $m_A$ and
$m_H$ become further lower. As the result,
both spin-independent scattering cross sections $\sigma_{\chi p}^{scalar}$ and 
spin-dependent one $\sigma_{\chi p}^{spin}$ are enhanced compared to mSUGRA 
case. 

In Fig.~\ref{fig:sigsip-sigsdp-tb35}, we present 
$\sigma^{scalar}_{\chi p}$ vs. $\sigma^{spin}_{\chi p} /(2 m_\chi/GeV)$  in
the NUHM (black points) and mSUGRA scenario (green points) 
for (a) $\tan\beta=35$ and (b) $\tan\beta=50$ respectively.
The dashed straight line in the figure indicates the region in which 
the two contributions to the capture rate are similar to each other. 
We observe that both $\sigma_{\chi p}^{spin}$ and especially 
$\sigma_{\chi p}^{scalar}$  in NUHM are enhanced a lot compared to 
mSUGRA scenario. An important point we notice from the figure is that 
$\sigma^{scalar}_{\chi p}$ is
usually (especially in the region of large cross section) larger than 
$\sigma^{spin}_{\chi p} /(2 m_\chi/GeV)$  in the NUHM case,  
while the opposite is true  for the mSUGRA case. 
This fact implies the capture rate for the sun in the NUHM 
is largely determined by the spin-independent scattering cross section 
rather than spin-dependent one, unlike the mSUGRA scenario. 
This is because the ratio of the contribution from spin-independent 
and spin-dependent cross section to the capture rate for the sun is 
approximately proportional to the ratio of 
$\sigma^{scalar}_{\chi p}$ and $\sigma^{spin}_{\chi p} /(2 m_\chi/GeV)$. 

\begin{figure}[ht!]
\begin{center}
\subfigure[]{
\includegraphics[height=7cm,width=7cm]{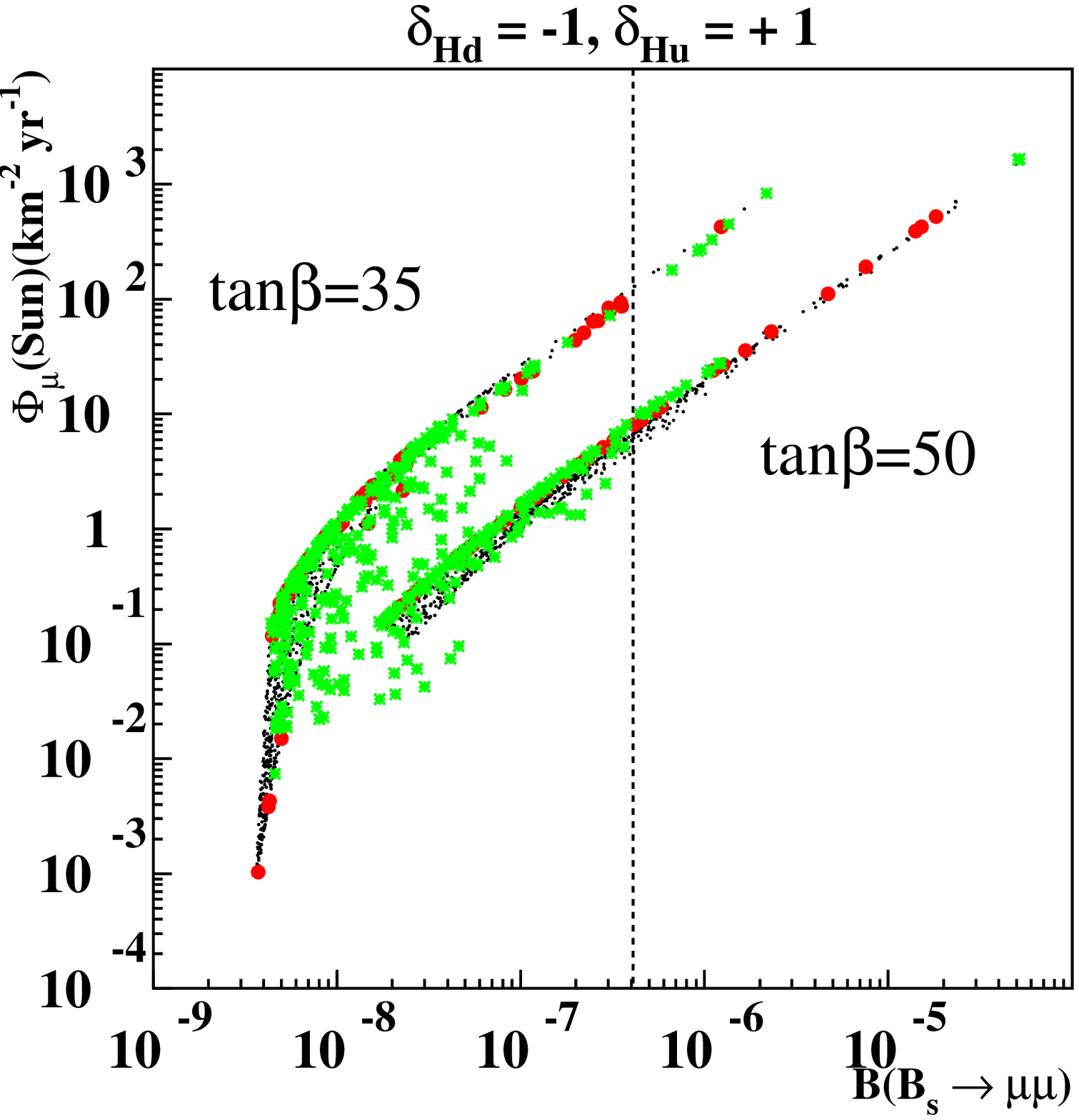}}\hskip 0.5cm
\subfigure[]{
\includegraphics[height=7cm,width=7cm]{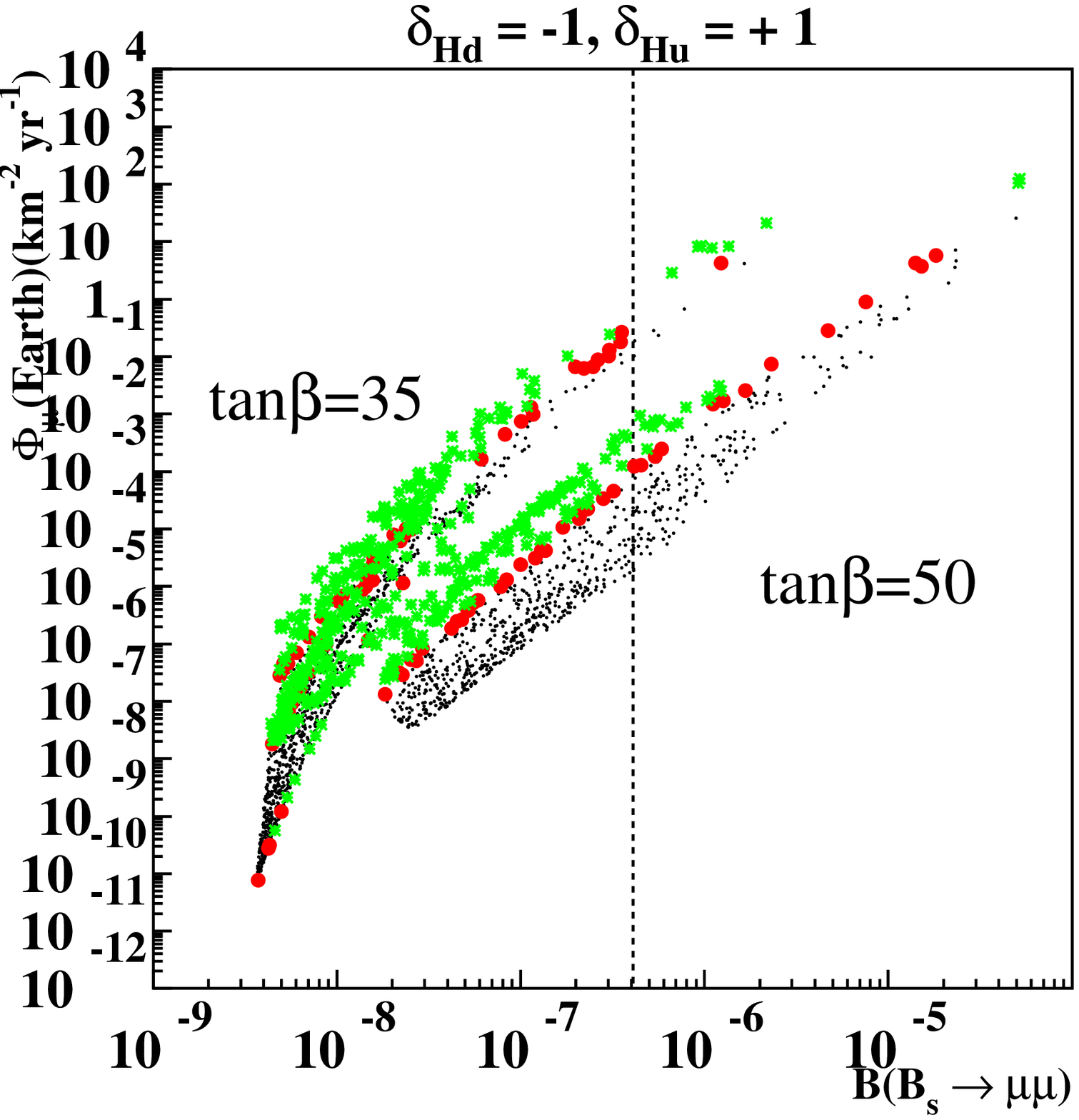}}
\end{center}
\vskip -0.5cm
\caption{ the muon fluxes from (a) the sun and (b) the earth 
vs. $B(B_s \rightarrow \mu^+ \mu^-)$
in Non-universal Higgs mass scenario.}
\label{fig:nuhm-bsmm}
\end{figure}

As we have shown in the previous paper \cite{bsdm1}, the current 
experimental limit  of $B(B_s \rightarrow \mu^+ \mu^-)$ puts a strong 
constraint on the allowed range of the spin-independent cross section.
Since the muon flux from the sun and the earth 
strongly depends on the spin-independent cross section, 
we naturally expect that the current limit of 
$B(B_s \rightarrow \mu^+ \mu^-)$ play 
an important part in restricting the muon flux. 
This point can be observed clearly in Fig.~\ref{fig:nuhm-bsmm},
where we  show explicitly the correlation between
$B(B_s \rightarrow \mu^+ \mu^-)$ and the muon flux from the sun (a)
and the earth (b) in NUHM for $\tan\beta=35$ and 50.
Note that the $B_s \rightarrow \mu^+ \mu^-$ is stronger for larger 
$\tan\beta$, and the resulting muon flux becomes smaller for the larger
$\tan\beta$ case, like the spin independent DM scattering cross section
\cite{bsdm1}.

\begin{figure}[ht!]
\begin{center}
\subfigure[]{
\includegraphics[height=7cm,width=7cm]{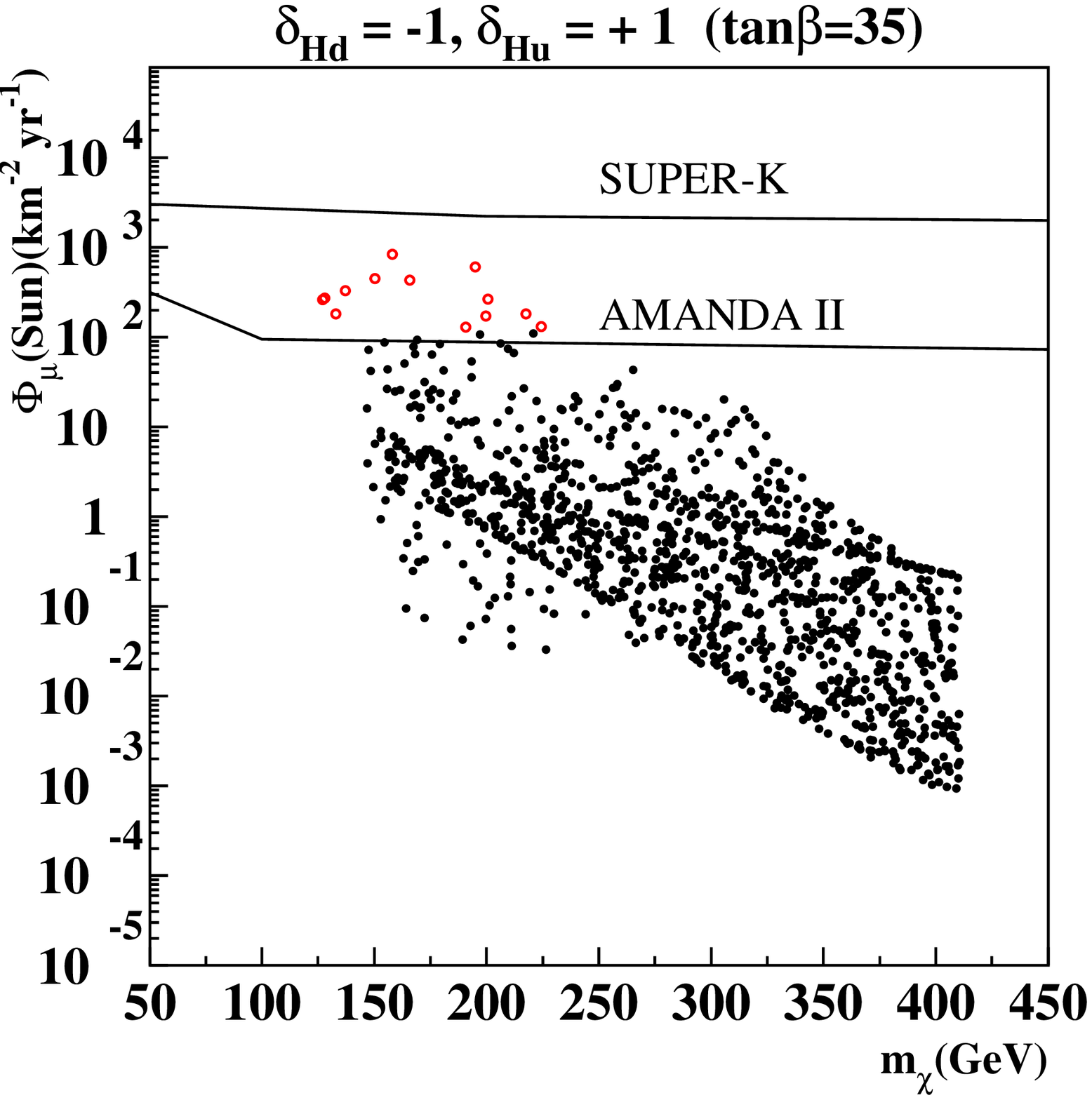}}\hskip 0.5cm
\subfigure[]{
\includegraphics[height=7cm,width=7cm]{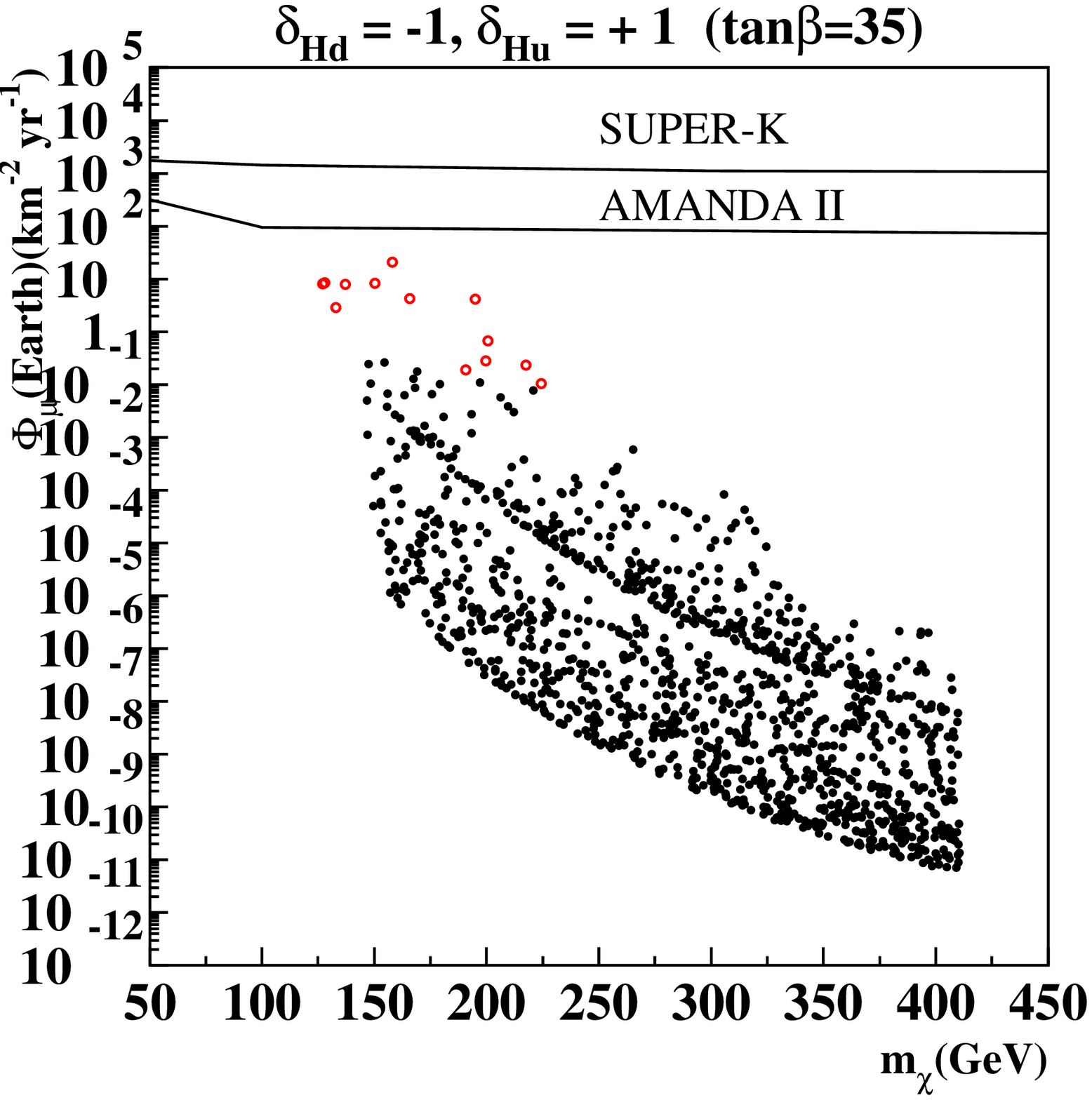}}
%
\subfigure[]{
\includegraphics[height=7cm,width=7cm]{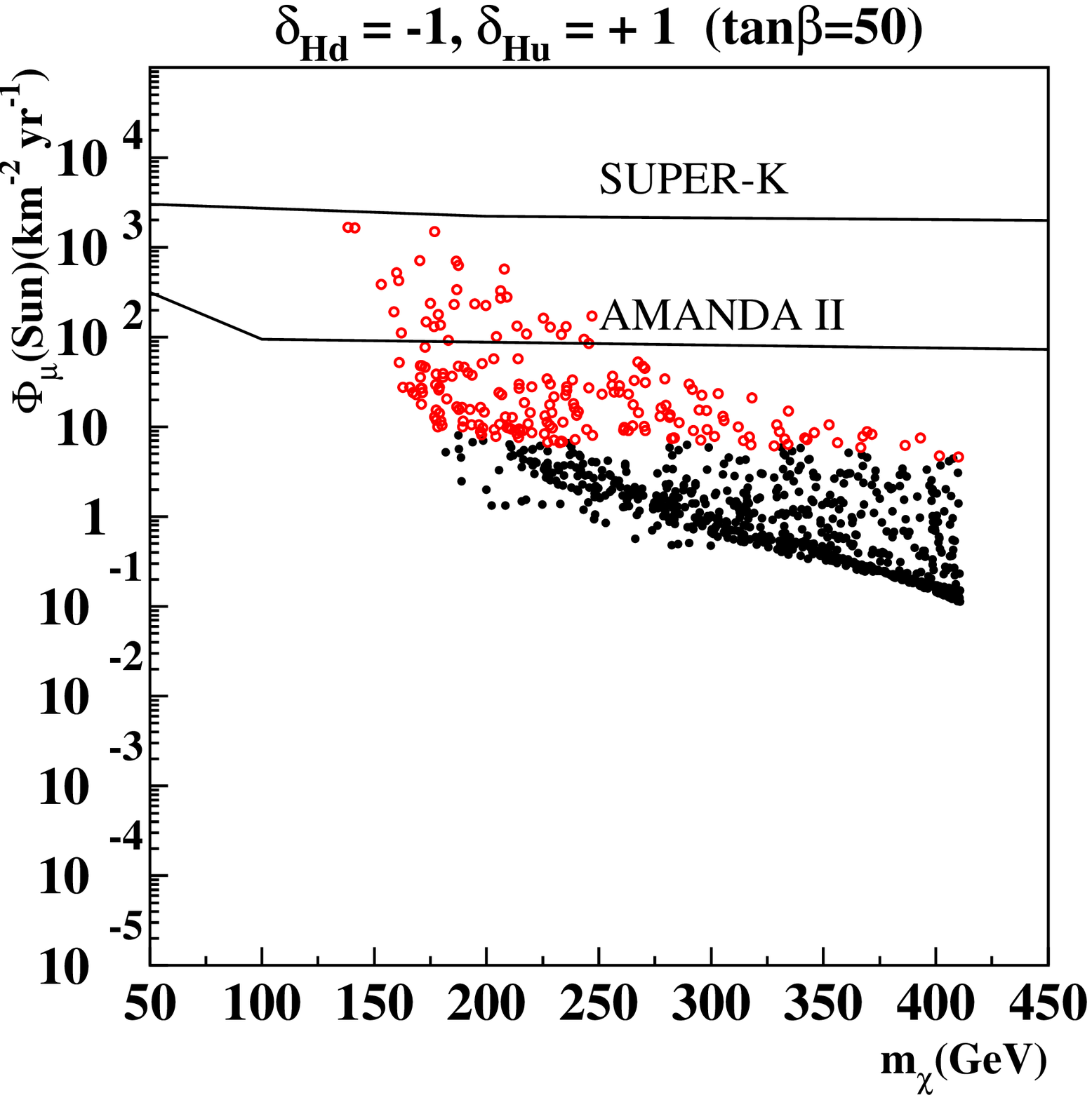}}\hskip 0.5cm
\subfigure[]{
\includegraphics[height=7cm,width=7cm]{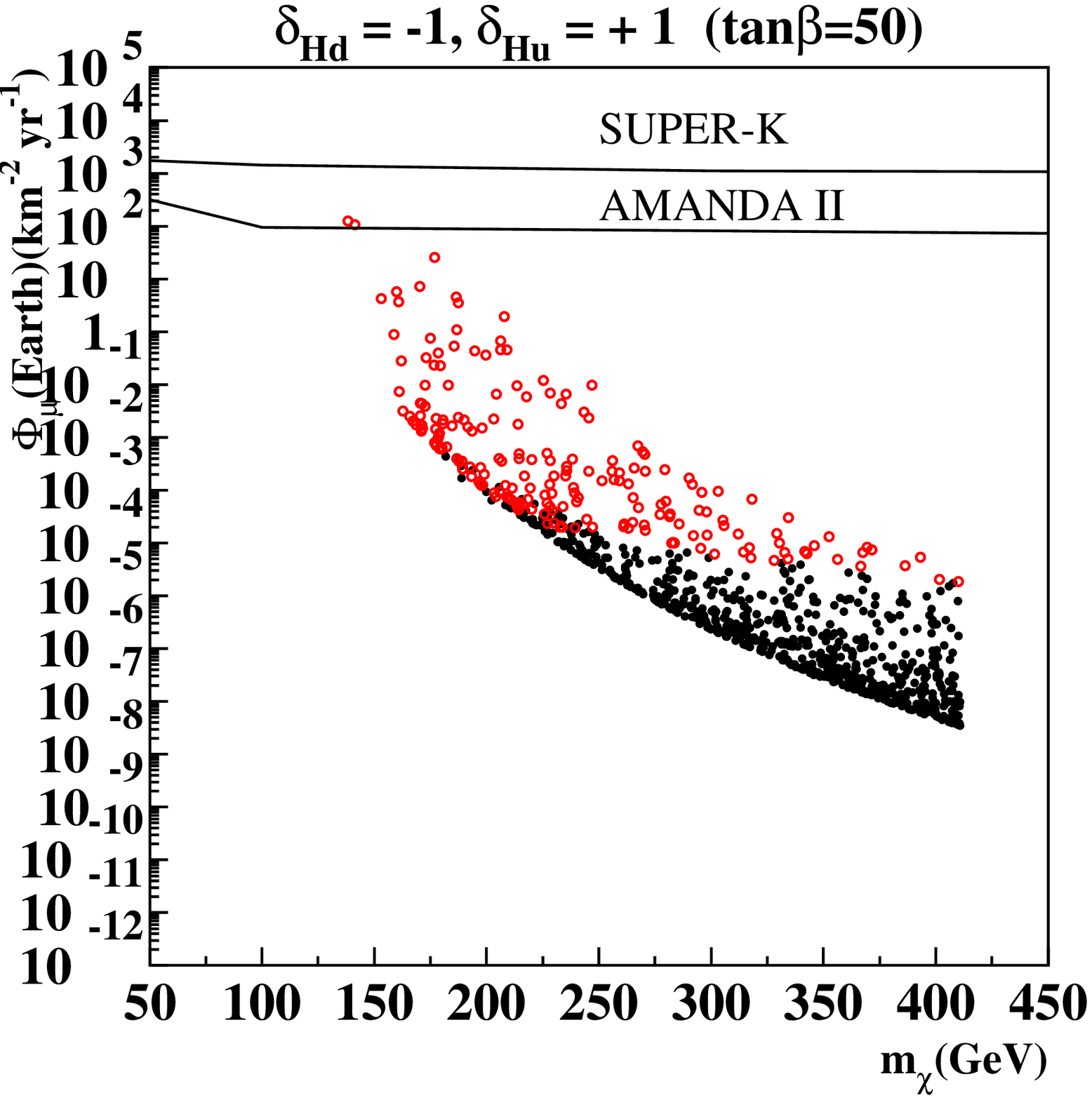}}
\end{center}
\vskip -0.5cm
\caption{ the muon flux from the sun and the earth 
vs. $m_\chi$ in Non-universal Higgs mass scenarios 
with $\delta_{H_d}=-1,\delta_{H_u}=+1$ and 
$\tan\beta = 35$ and $\tan\beta=50$. 
The red points (the open circles) are exclued by the current upper limit of 
$B(B_s \rightarrow \mu^+ \mu^-)$.}
\label{fig:nuhm-tb50}
\end{figure}

The enhancements of the neutralino DM scattering cross sections (both 
spin-dependent and spin-independent) lead to 
the substantial change of the muon flux both from the sun and the earth 
compared with the mSUGRA case.
Fig.s~\ref{fig:nuhm-tb50} (a) and (b) show the muon fluxes from the 
sun and the earth, respectively,  in non-universal Higgs mass scenario 
with $\delta_{H_d}=-1, \delta_{H_u}=+1$  for $\tan\beta=35$ case.
Now the maximal values of the muon fluxes from the sun and the earth are 
$\sim 10^3~(10)~km^{-2} yr^{-1}$, which is two (eight) orders of magnitude 
lager than  the one for the mSUGRA case with $\tan\beta=35$.
In Fig.~\ref{fig:nuhm-tb50} (c) and (d), we show the muon flux 
from the sun and the earth, respectively, in 
non-universal Higgs mass scenario with $\delta_{H_d}=-1, \delta_{H_u}=+1$ 
for $\tan\beta=50$.
The red points (the open circles) are exclued by the current upper limit of 
$B(B_s \rightarrow \mu^+ \mu^-)$.

\subsection{D-brane model}

Next, we consider a specific $D$ brane model where the SM gauge groups
and 3 generations live on different $Dp$ branes \cite{dbrane}.
In this model, scalar fermion masses are not completely universal and
gaugino mass unification can be relaxed. Also the string scale is around
$10^{12}$ GeV (the intermediate scale) rather than GUT scale.

Since there are now three moduli ($T_i$)  and one dilaton superfields
in this case, we use the following parametrization that is appropriate for
several $T_i$ moduli:
\begin{eqnarray}
  F^S & = & \sqrt{3}~( S + S^* )~ m_{3/2} \sin\theta,
\nonumber   \\
  F^i & = & \sqrt{3}~( T_i + T_i^* )~ m_{3/2} \cos\theta~\Theta_i
\end{eqnarray}
where $\theta$ and $\Theta_i ~(i = 1,2,3)$ with $\sum_i | \Theta_i |^2 = 1$
parametrize the directions of the goldstinos in the $S, T_i$ field space.
Then, the gaugino masses are given by
\begin{eqnarray}
M_3 & = & \sqrt{3} m_{3/2} \sin\theta ,
\nonumber   \\
M_2 & = & \sqrt{3} m_{3/2} \Theta_1 \cos\theta ,
\nonumber   \\
M_Y & = & \sqrt{3} m_{3/2} \alpha_Y ( M_I ) ~\left( { 2 \Theta_3 \cos\theta
\over \alpha_1 ( M_I ) } + {\Theta_1 \cos\theta \over \alpha_2 ( M_I )}
+ {2 \sin\theta \over 3 \alpha_3 ( M_I ) } \right) ,
\end{eqnarray}
where
\begin{equation}
  {1\over \alpha_Y ( M_I ) } = { 2 \over \alpha_1 ( M_I ) }
+ { 1 \over \alpha_2 ( M_I ) } + {2 \over 3 \alpha_3 ( M_I ) }.
\end{equation}
The string scale $M_I$ is determined to be
$M_I = 10^{12} ~(5 \times 10^{14} )$ GeV from the $U(1)_1$ gauge coupling
$\alpha_1 ( M_I ) = 0.1 (1)$ \cite{dbrane}.  
Note that the gaugino masses are non universal in a natural way in this 
scenario, unlike other scenarios studied in the previous subsections.

The soft masses for the sfermions and Higgs fields are given by
\begin{eqnarray}
  m_Q^2     & = & m_{3/2}^2~ \left[~ 1 - {3\over 2} \left( 1 - \Theta_1^2
\right)~ \cos^2 \theta~  \right],
\nonumber   \\
  m_{u^c}^2 & = & m_{3/2}^2~ \left[~ 1 - {3\over 2} \left( 1 - \Theta_3^2
\right)~ \cos^2 \theta~  \right],
\nonumber   \\
  m_{d^c}^2 & = & m_{3/2}^2~ \left[~ 1 - {3\over 2} \left( 1 - \Theta_2^2
\right)~ \cos^2 \theta~  \right],
\nonumber   \\
  m_L^2     & = & m_{3/2}^2~ \left[~ 1 - {3\over 2} \left( \sin^2 \theta
+ \Theta_3^2 ~\cos^2 \theta \right) \right],
\nonumber   \\
  m_{e^c}^2 & = & m_{3/2}^2~ \left[~ 1 - {3\over 2} \left( \sin^2 \theta
+ \Theta_1^2 ~\cos^2 \theta \right) \right],
\nonumber   \\
  m_{H_2}^2 & = &  m_{3/2}^2~ \left[~ 1 - {3\over 2} \left( \sin^2 \theta
+ \Theta_2^2 ~\cos^2 \theta \right) \right],
\nonumber   \\
  m_{H_1}^2 & = & m_L^2.
\end{eqnarray}
Note that the scalar mass universality in the sfermion masses and Higgs
masses is achieved when
\begin{equation}
\sin^2 \theta = {1\over 4} ~~~{\rm and}~~~ \Theta_i^2 = {1 \over 3}~~~
{\rm for}~i = 1,2,3.
\end{equation}
And in this case the gaugino masses becomes also universal, when we take
only positive numbers for the solutions. For other choices of goldstino
angles, the scalar and the gaugino masses become nonuniversal, and there
could be larger or smaller flavor violation in the low energy processes 
as well as enhanced SUSY contributions to the $a_\mu^{\rm SUSY}$. 

The trilinear couplings are given by
\begin{eqnarray}
  A_u & = & {\sqrt{3} \over 2}~m_{3/2}~\left[
( \Theta_2 - \Theta_1 - \Theta_3 ) \cos\theta - \sin\theta \right],
\nonumber   \\
  A_d & = & {\sqrt{3} \over 2}~m_{3/2}~\left[
( \Theta_3 - \Theta_1 - \Theta_2 ) \cos\theta - \sin\theta \right],
\nonumber   \\
  A_e & = & 0.
\end{eqnarray}
Therefore the $D$ brane model considered in this work is specified by
following six parameters :
\[
m_{3/2},~~\tan\beta,~~\theta,~~\Theta_{i=1,2},~~ {\rm sign} (\mu).
\]

\begin{figure}[ht!]
\begin{center}
\subfigure[]{
\includegraphics[height=7cm,width=7cm]{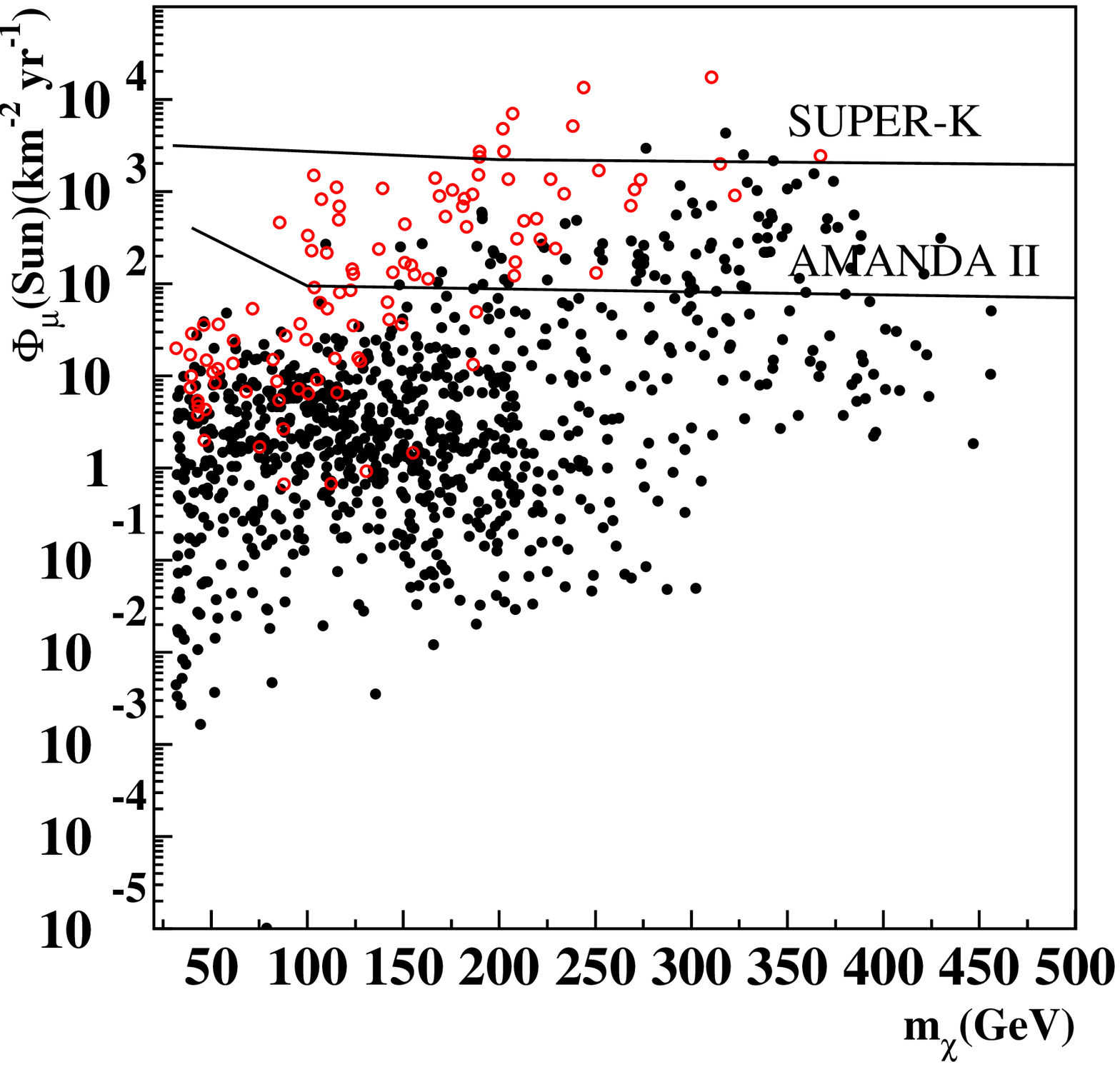}}\hskip 0.5cm
\subfigure[]{
\includegraphics[height=7cm,width=7cm]{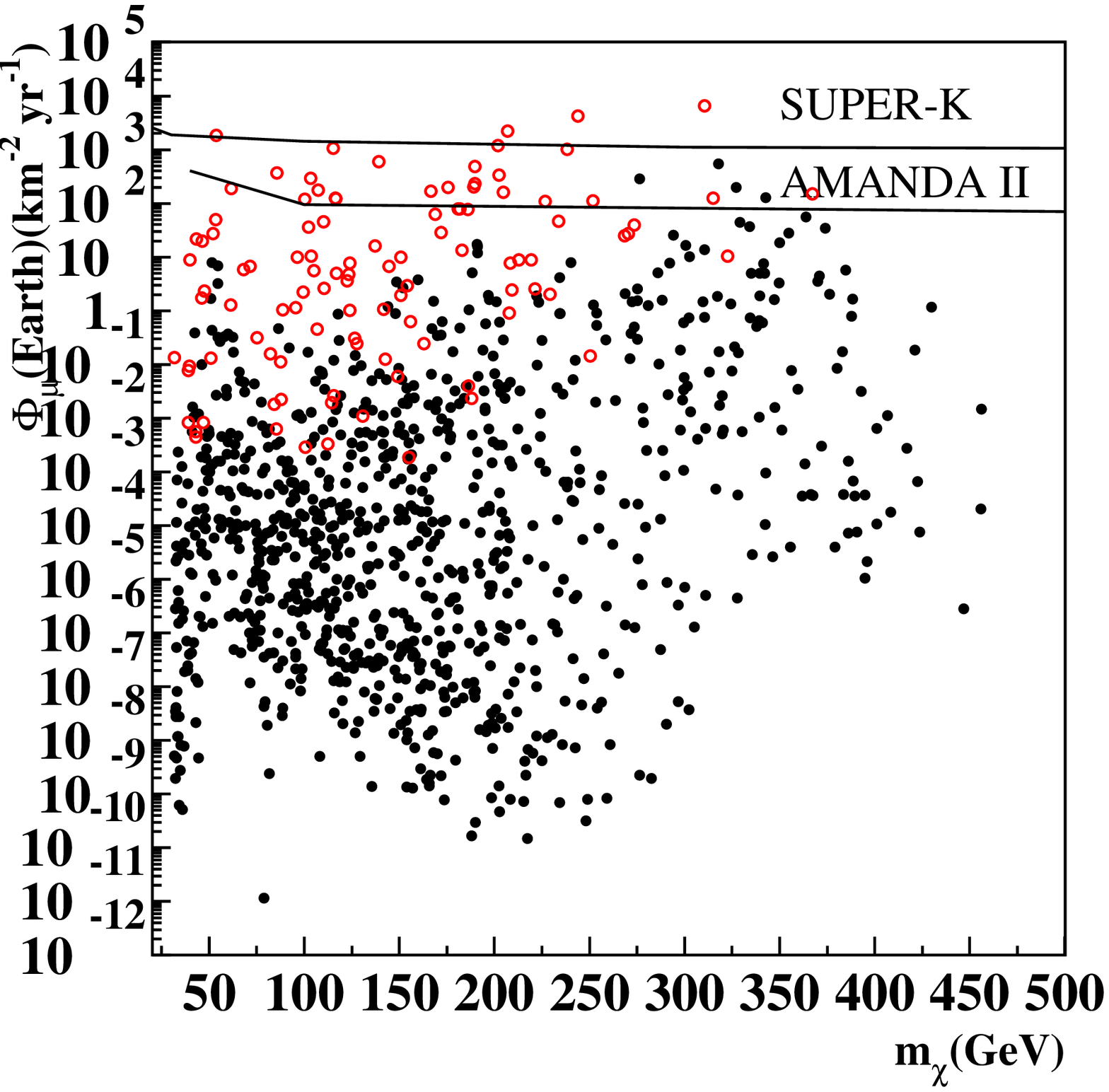}}
\end{center}
\vskip -0.5cm
\caption{ the muon flux from the sun  and the earth 
vs. $m_\chi$ in a D-brane model with $\tan\beta=50$. 
The red points (the open circles) are exclued by the current upper limit of 
$B(B_s \rightarrow \mu^+ \mu^-)$.}
\label{fig:dalitz}
\end{figure}

Due to the departure from the universlity of scalar masses and the 
proportionality of trilinear couplings, the flavor violation could be
different from the mSUGRA case. For example, it is possible to have smaller
$b\rightarrow s$ transition due to the smaller  $\tilde{t}_L - \tilde{t}_R$
mixing and larger stop masses in this $D-$brane scenarios, so that the 
$B ( B_s \rightarrow \mu^+ \mu^- )$ constraint can be relaxed. 
This can be seen in Fig.~\ref{fig:dalitz} (a) and (b), where
the large flux signals are excluded by Super-K and AMANDA II, but not 
by the $B ( B_s \rightarrow \mu^+ \mu^- )$ constraint. 
In this limited parameter space, one can have a large DM scattering
cross section and the upward-going muon flux without conflict with the
$B \rightarrow \mu^+ \mu^-$ branching ratio. Also there is no strong
correlations among these observables.  Therefore the indirect
search for the DM annihilation is complementary to the 
$B_s \rightarrow \mu^+ \mu^-$ branching ratio in the $D-$brane scenarios.  

\section{Conclusions}
\label{sec:conclusion}

In this work, we considered the indirect detection of the DM through the
upward-going muon flux from the DM annihilation at the core of the sun or
the earth, along with the upper bound on the branching ratio for the 
$B_s \rightarrow \mu^+ \mu^-$ decay, in some general supergravity scenarios
where the upward-going muon flux could be enhanced very much compared to the
mSUGRA case. 
In general supergravity scenario with non-universal Higgs model, we found
the following: 
\begin{itemize}
\item Both $\sigma_{\chi p}^{spin}$ and $\sigma_{\chi p}^{scalar}$ can be 
enhanced a lot compared to the mSUGRA scenario, but the enhancement in
the spin-independent part is much greater. 
\item Therefore, contrary to the usual claim, the upward-going muon  
flux from the sun can be  dominated by the  spin-independent  part 
$\sigma_{\chi p}^{scalar}$ in the NUHM, rahter than by the spin-dependent part 
$\sigma_{\chi p}^{spin}$, as in the mSUGRA scenario [ Fig.~2 (a) and (b) ].
\item The current upper bound 
$B ( B_s \rightarrow \mu^+ \mu^- ) < 4.1 \times 10^{-7}$ excludes a large
parameter space where the muon fluxes could be enhanced otherwise, and
the constraint is stronger for larger $\tan\beta$ [ Fig.~3 (a) and (b) ]. 
\item The upper bound on $B ( B_s \rightarrow \mu^+ \mu^- )$ becomes
much stronger than the upper limits on the muon flux from Super-K and 
AMANDA II  [ Fig.~4 (a)--(d) ].
\end{itemize} 

In the $D-$brane models with nonuniversal scalar fermion masses, the
correlations between the muon flux and $B ( B_s \rightarrow \mu^+ \mu^- )$ 
becomes lost, and the upper bound on  $B ( B_s \rightarrow \mu^+ \mu^- )$ 
is complementary to the upper bounds on the muon fluxes from Super-K
and AMANDA II. 
Our study shows that the muon flux originated from the DM annihilation
in the sun could be in the range of a few $\times 10^3 $ /km$^2 \cdot$ yr.

Our study indicates that it is most important to include the 
$B_s \rightarrow \mu^+ \mu^-$ branching ratio constraint   
when we study the direct and the indirect detections 
of the neutralino DM in general supergravity scenarios. 
The upper limit on the $B_s \rightarrow \mu^+ \mu^-$ branching ratio 
excludes significant part of parameter space where the DM scattering cross
section and the upward-going muon flux could be enhanced above/around
the current experiments. Unless the chargino-stop contribution 
to $B_s \rightarrow \mu^+ \mu^-$
is very small or there is fortuitous cancellation between the chargino-stop
and the gluino-sbottom loop contributions,
the spin-independent DM scattering cross section and the indirect detection
rate through the upward-going muon flux are strongly contrained by 
the $B_s \rightarrow \mu^+ \mu^-$ branching ratio. 
Since both the direct and the indirect detection rates are well below 
the current experiments in most supergravity model parameter space when
the $B_s \rightarrow \mu^+ \mu^-$ branching ratio constraint is imposed,
it would be a great challenge for experimemtalists to reach such 
sensitivity 
to have positive signals of the DM search.

\bigskip

\acknowledgments
The authors are grateful to A. Masiero and P. Ullio for useful discussions on indirect
detection of the neutralino DM.  
PK is supported in part by KOSEF Sundo grant R02-2003-000-10085-0,
KOSEF through CHEP at Kyungpook National  University and  
KRF grant KRF-2002-070-C00022. 
The work of YGK was supported by Korea Research Foundation and 
the Korean Federation of Science and Technology Societies Grant funded 
by Korea Government (MOEHRD, Basic Research Promotion Fund).



\listoftables           
\listoffigures          

\end{document}